\documentclass[prd,showpacs,preprintnumbers,floatfix,superscriptaddress,10pt]{revtex4}
\usepackage[mathscr]{eucal}
\usepackage{color}
\usepackage[dvips]{graphicx}
\usepackage{epsf}
\usepackage{bm}
\usepackage{epsfig}
\usepackage{feynmf}
\usepackage{amsmath}

\newcommand{\be}{\begin{equation}}
\newcommand{\ee}{\end{equation}}
\newcommand{\ba}{\begin{eqnarray}}
\newcommand{\ea}{\end{eqnarray}}

\newcommand{\tint}{\int \frac{d^3q}{(2\pi)^3} }

\begin{document}

\title{Effective field theory  and dispersion law of the phonons of a non-relativistic superfluid}

\author{Miguel Angel Escobedo}
\email{mesco@ecm.ub.es}
\affiliation{
Departament d'Estructura i Constituents de la Mat\`eria and \\
Institut de Ci\`encies del Cosmos, Universitat de Barcelona \\
Diagonal 647, E-08028 Barcelona, Catalonia, Spain }

\author{Cristina Manuel}
\email{cmanuel@ieec.uab.es}
\affiliation{Instituto de Ciencias del Espacio (IEEC/CSIC) Campus Universitat Aut\`onoma de Barcelona, Facultat de Ci\`encies, Torre C5, E-08193 Bellaterra (Barcelona), Catalonia, Spain}
\date{19th April 2010}
\preprint{UB-ECM-PF-10/11}
\begin{abstract}
We study the recently proposed effective-field theory for the phonon
of an arbitrary nonrelativistic superfluid. After computing the one-loop phonon self-energy,
we obtain the low-temperature $T$ contributions to the phonon dispersion law
at low momentum and see that the real part of those can be parametrized
as a thermal correction to the phonon velocity.
 Because the phonons are the quanta of the sound waves, at low momentum their
velocity should agree with the speed of sound.  We find that our results match at order $T^4 \ln{T}$ with those predicted 
by Andreev
and Khalatnikov for the speed of sound, derived from the superfluid hydrodynamical equations
and the phonon kinetic theory.
We get also higher-order corrections of order $T^4$, which are not
reproduced pushing naively the kinetic theory computation.
Finally, as an application, we consider  the cold Fermi gas in the unitarity limit and 
find a universal expression for the   low-$T$ relative correction to the speed of sound for these systems.
\end{abstract}

\maketitle

\section{Introduction}


Superfluidity is a phenomenon that occurs at low temperatures
after the appearance of a quantum condensate, in either bosonic or fermionic systems  ~\cite{landaufluids,IntroSupe}.
The condensate spontaneously breaks the global $U(1)$ symmetry associated with particle number conservation
of the system. In such a case  Goldstone's theorem predicts the existence of low-energy modes that at sufficiently low momentum have  a linear dispersion law and are essential to explain the property of superfluidity. 
We refer generically to these modes as superfluid phonons, or phonons for simplicity.

The phonons dominate the physics in the superfluid at long wavelengths. At very low temperatures they also dominate the thermal corrections to the
thermodynamical and hydrodynamical properties of the superfluid. It is then important to have a precise knowledge of its self-interactions to know with  accuracy the dynamics of these systems.
Landau gave a successful phenomenological description of superfluidity and derived the main
phonon self-interactions by imposing an {\it ad hoc} phonon Hamiltonian at leading order
valid for $^4$He. Later on, it was shown that the energy spectrum for a weakly interacting Bose gas, which experiences a Bose-Einstein condensation, could be derived \cite{IntroSupe,Griffin}. The transport equation obeyed
by the phonons can also be derived for the weakly interacting Bose gas \cite{Kirkpatrick}.

Effective-field theory (EFT) techniques have been proposed  and used with success in particle
physics, although they have not been  explored much in other branches of physics \cite{Weinberg,Burgess:1998ku}. The techniques are
especially suited for systems where there is a hierarchy of widely separated energy or momentum scales.
Then, one performs an expansion in powers of energy or momentum, rather than expanding
in a coupling constant. Power counting and symmetry considerations fix the form of the
EFT to the accuracy one desires, regardless of whether the underlying system
is weakly or strongly coupled. If the theory describes only the low-energy degrees of freedom,
the dynamics of the short length physics is encoded in the value of the coupling constants of
the effective theory. One remarkable successful example of the application of EFT techniques is chiral
perturbation theory to describe the long-wavelength physics of the nuclear strong interactions \cite{Gasser:1983yg}.

EFT techniques have been recently applied to study superfluid systems.
The phonon Lagrangian at leading order in the momentum expansion and also the phonon Lagrangian at next-to-leading
order have been explicitly derived, with an emphasis on applications to the superfluid regime
of the cold unitary Fermi gas \cite{Son:2002zn,Andersen:2002nd,Son:2005rv,valle,Stachel}. These are systems where the strength of the fermion interactions,
measured in terms of the $s$-wave scattering length, is asymptotically large (see Ref.~\cite{Giorgini:2008zz} for a review). In this limit those
systems exhibit conformal invariance, and they are believed to have universal properties \cite{Ho:2004zza}, meaning
that their features do not depend on the detailed form of the interparticle potential.
EFT techniques might be very conveniently applied to the superfluid phase of the cold unitary Fermi gas, as
these are strongly coupled.

This article is focused on the study of the recently proposed EFT to describe the phonons
of a nonrelativistic superfluid and to extract from it their dispersion law.
We discuss some  subtleties that appear in the computations like, due to the fact that
the phonons are massless degrees of freedom,  the traditional EFT power counting gets mixed up when
going close to the phonon on-shell limit. This problem shows up when studying the phonon dispersion law,
as we discuss at length.
We then focus on  getting thermal corrections to the phonon dispersion law. 

We only consider the very low-temperature $T$ regime of the superfluid, where the phonons
are in a collisionless or nonhydrodynamical domain. In this situation, the phonons can be viewed
as forming a bosonic gas, rather than a fluid. The hydrodynamical behavior of the superfluid in this
situation simplifies, as opposed to what happens at higher $T$ where the description of the system
requires the two-fluid model of Landau. At $T=0$, or  very low $T$, there is only one fluid in the system,
and one can talk about one
 single sound hydrodynamical velocity.

At zero temperature, and very low momentum, the phonons obey
\begin{equation}
\label{lin-dlaw}
\lim_{p\to 0} E_p =c_s p \ ,
\end{equation}
where $c_s$ is the velocity of sound. This relation was suggested by Landau for the superfluid regime of $^4$He,
as he realized that the elementary excitations in the system were the quanta of the sound waves \cite{IntroSupe}.
The relation has been proven for a Bose superfluid by Gavoret and Nori\`eres in Ref.~\cite{Gavoret} for $T=0$. 
We are not aware of the existence of a general proof that the
same relation should hold at very-low-temperature $T$.  However, because the phonons are the quanta of the sound waves,
 it is  expected that for very low momenta their properties should be the same as those of the hydrodynamical sound waves
 of the system.

The speed of sound receives thermal corrections, first computed
 by Andreev and Khalatnikov \cite{Andreev} in the collisionless regime of $^4$He,
using both Landau's hydrodynamical equations and the phonon kinetic theory of Khalatnikov.
This computation was believed to be valid for other superfluids as well.
Thermal corrections to the phonon velocity have only been computed for the
weakly interacting Bose gas \cite{Singh,Ma}, reaching agreement with the speed of sound correction
in the  terms of order $T^4 \ln{T}$ (see also Ref.~\cite{Giorgini}).
In this article we  review the Andreev and Khalatnikov \cite{Andreev} computation of the thermal
corrections to the speed of sound and extend it to get the subleading corrections in $T$.
We compare such a computation with the thermal corrections to the phonon velocity in the
low-momentum limit, as derived from the EFT. We  find agreement in the two quantities for the
leading corrections that go as $T^4 \ln{T}$, but not for the subleading corrections of order $T^4$.
We discuss the possible origin of these discrepancies.

This article is organized as follows. In Sec.~\ref{phononEFT}
we review the phonon EFT  at leading order and next-to-leading order in
a derivative expansion \cite{Son:2005rv}. We present the analysis
of the one-loop phonon self-energy with the leading-order Lagrangian in
Sec.~\ref{1loopSec}, both at $T=0$ (Sec.~\ref {T0-oneloop})
 and finite $T$ (Sec.~\ref{T-oneloopcor}),
 and discuss the need to go to the next-to-leading
order for the computation of the phonon dispersion law in Sec.~\ref{1loopNLO}.
The phonon dispersion law derived from all the one-loop corrections is 
presented in Sec.~\ref{phdisp-sec}. In Sec.~\ref{kinetic+hydro}, we review the
computation of the thermal corrections to the speed of sound by Andreev and Khalatnikov
and push it to get subleading corrections in $T$. In Sec.~\ref{coldFgas-sec} we concentrate on getting the thermal
corrections to the speed of the phonon of the cold Fermi gas in the unitarity limit.  Our conclusions are presented in
Sec.~\ref{conclu}. Appendix~\ref{set-relations} is devoted to showing some useful relations among
 the self-coupling constants of the phonon EFT, and in Appendix~\ref{App-LOthermal}
 we show explicit details of the computation of the thermal corrections to the phonon
 self-energy. Throughout the article we use natural units, $\hbar = k_B =1$.

\section{The superfluid phonon effective-field theory at leading  and next-to-leading orders}
\label{phononEFT}


The superfluid phonon is the Goldstone mode associated with the spontaneous symmetry breaking
of a $U(1)$ symmetry, which corresponds to particle number conservation.
EFT techniques can be used to write down the effective Lagrangian associated with the superfluid phonon.  The effective Lagrangian
is then presented as an expansion in derivatives of the Goldstone field, the terms of this
expansion being 
restricted by symmetry considerations.
 The coefficients of the Lagrangian can be, in principle, computed from the microscopic theory through a standard
matching procedure, and thus  they depend on the short-range physics of the system under
consideration.

It has been known for a while that the leading-order (LO) term Lagrangian of the Goldstone
mode of a superfluid system is entirely fixed by  the equation of state~\cite{Popov,Greiter:1989qb,Liu:1998ey}. 
In some more recent publications \cite{Son:2002zn,Son:2005rv} it has been realized that at the
lowest order in a derivative expansion the Lagrangian reads (see Sec.~V of Ref.~\cite{Son:2005rv})
\begin{equation}
\label{LO-Lagran}
\mathcal{L}_{\rm LO} =P (X) \ ,
\end{equation}
where $P(\mu_0)$ and $\mu_0$ are the pressure and chemical potential, respectively, of the superfluid at $T=0$, and
\begin{equation}
 X = \mu_0-\partial_t\varphi-\frac{({\bf \nabla}\varphi)^2}{2m} \ ,
\end{equation}
where $\varphi$ is the phonon field, and $m$ is the mass of the
particles that condense.  The form of $X$ is dictated by imposing Galilean invariance to the EFT.
In Ref.~\cite{Son:2005rv} it was noted that after a Legendre
transform, one can associate this formulation with Popov's formulation \cite{Popov}.

The final reason why the LO Lagrangian takes this particular form is that the effective action
associated with the theory at its minimum for constant classical field configurations has to
be equal to the pressure \cite{Son:2002zn}. This formulation turns out to be  very advantageous,
as it allows one to derive all the phonon properties at the lowest order in momentum based on the
 knowledge of the zero temperature pressure of the superfluid. In particular, one can easily
get the phonon dispersion law
and the form of the leading phonon self-interactions, and in principle their leading contribution to different physical
processes.

 In order to see this, we expand the
Lagrangian around $P(\mu_0)$ in a Taylor series, and after the field redefinition
\begin{equation}
\varphi = \frac{\phi}{\sqrt{\frac{\partial^2P}{\partial\mu_0^2}}} \ ,
\end{equation}
to have a canonically normalized kinetic term, one can write
\begin{equation}
\label{comlag}
\mathcal{L}_{\rm LO}=\frac{1}{2}\left((\partial_t\phi)^2-v^2_{\rm ph}({\bf \nabla}\phi)^2\right)-g\left((\partial_t \phi)^3-3\eta_g \,\partial_t \phi({\bf \nabla}\phi)^2 \right)
+\lambda\left((\partial_t\phi)^4-\eta_{\lambda,1} (\partial_t\phi)^2({\bf \nabla}\phi)^2+\eta_{\lambda, 2}({\bf \nabla}\phi)^4\right)
+ \cdots
\end{equation}
We have neglected an irrelevant constant and a total time derivative term,
which is only needed to study vortex configurations.

All the coefficients that appear in Eq.~(\ref{comlag}) can be expressed in terms of different ratios of derivatives of the pressure.
In particular, after using the thermodynamical relation $dP = \frac{\rho_0}{m} d\mu_0$ ,
where  $\rho_0$ is the mass density at $T=0$, one can check that at this
order the phonon velocity is
\begin{equation}
v_{\rm ph}= \sqrt{\frac{\frac{\partial P}{\partial\mu_0}}{m\frac{\partial^2P}{\partial\mu_0^2}}} =  \sqrt{\frac{\partial P}{\partial\rho_0}}  \equiv c_s \ ,
\end{equation}
that is, it  can be identified with the speed of sound at $T=0$, as is
expected in the low-momentum limit. The dispersion law obtained from this Lagrangian, neglecting both 
quantum and thermal corrections,   is exactly $E_p = c_s p$.

The remaining coefficients of Eq.~(\ref{comlag}) give account of the phonon self-couplings and are expressed
as
\begin{equation}
\label{coup-constants}
g =\frac{\frac{\partial^3P}{\partial\mu_0^3}}{6\left(\frac{\partial^2P}{\partial\mu_0^2}\right)^{3/2}} \ , \qquad
\eta_g= \frac{\frac{\partial^2P}{\partial\mu_0^2}}{m\frac{\partial^3P}{\partial\mu_0^3}} \ , \qquad
\lambda = \frac{\frac{\partial^4P}{\partial\mu_0^4}}{24\left(\frac{\partial^2P}{\partial\mu_0^2}\right)^2} \ , \qquad
\eta_{\lambda,1} =\frac{6\frac{\partial^3P}{\partial\mu_0^3}}{m\frac{\partial^4P}{\partial\mu_0^4}} \ , \qquad
\eta_{\lambda,2} = \frac{3\frac{\partial^2P}{\partial\mu_0^2}}{m^2\frac{\partial^4P}{\partial\mu_0^4}}
\ .
\end{equation}

In the following section we study  the one-loop phonon self-energy, and we see
that different combinations of the aforementioned coupling constants appear.
For purposes of comparison with the kinetic theory approach, it is convenient
to express these combinations in terms of  derivatives of the speed of sound and of
$\rho_0$, rather than in terms of derivatives of the pressure.
 We present in Appendix~\ref{set-relations} some of the relations that are used later on.

The next-to-leading order  (NLO) Lagrangian in the derivative expansion has been constructed by Son and Wingate in
 Ref.~\cite{Son:2005rv}
(see also Ref.~\cite{valle}) by coupling the system to both external gauge $A_0$ and gravitational
fields and demanding both coordinate and gauge invariance. For $\theta = \mu_0 t - \varphi$, it
reads 
\begin{equation}
\label{NLO-Lagran}
\mathcal{L}_{\rm NLO} = \partial_i X  \partial_i X f_1(X) + \left(\nabla^2 \theta \right)^2 f_2(X)+
m \nabla^2 A_0 f_3(X) \ ,
\end{equation}
where $f_1, f_2$ and $f_3$ are arbitrary functions \cite{footnote}, and for simplicity we have assumed $A_0 =0$, as here we are not interested in the effect
 of an external potential in the phonon dynamics.
For superfluid systems in the  unitarity limit  
conformal invariance further restricts the form of  the extra functions $f_i (X)$ \cite{Son:2005rv}
and imposes $f_1(\mu_0) = c_1 (m/\mu_0)^{1/2}$ and $f_2(\mu_0) = c_2 (\mu_0/m)^{1/2}$, where $c_1$ and $c_2$ are dimensionless constants, which might be obtained from the microscopic theory by evaluating the
static density and transverse response functions \cite{Son:2005rv}. 
For the cold Fermi gas, these have been evaluated using the $\epsilon$ expansion in Ref.~\cite{Rupak:2008xq}.

When one works at the NLO, there are derivative corrections to the LO vertices, and also there
are new vertices which are not present at the LO. But most important for the purposes of this
article is that
the phonon dispersion relation is now of the form  
\begin{equation}
\label{displaw-NLO}
E_p = c_s p (1 - \gamma p^2) + \cdots
\end{equation}
 The parameter
$\gamma$ is a function of the coefficients that appear in the Lagrangian at the NLO:
\begin{equation}
\gamma=\frac{1}{\frac{\partial^2 P}{\partial\mu_0^2}}\left(f_1(\mu_0)+\frac{f_2(\mu_0)}{c_s^2}\right) \ .
\end{equation}
The sign of the parameter $\gamma$ is very important in the phonon physics. If $\gamma \leq 0$, then
the process of one phonon decaying into two is kinematically allowed, but it is not allowed in the opposite case.
This fact has several implications in the evaluation of the phonon damping, as we show later on.

 The parameter $\gamma$ for the cold Fermi gas has been obtained
from a phenomenological fit to numerical quantum Monte Carlo simulations in Ref.~\cite{salasnich}; the authors of this reference fitted to the Monte Carlo data of Ref.~\cite{Blume}.

\begin{figure}[!t]
\includegraphics[width=1.5in,angle=-0]{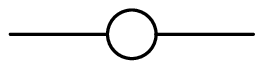}
\includegraphics[width=1.5in,angle=-0]{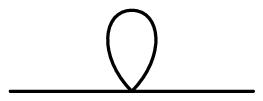}
\includegraphics[width=1.5in,angle=-0]{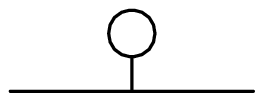}
\caption{One-loop Feynman diagrams for the phonon self-energy, dubbed as
 bubble, tadpole and lollypop  diagrams, respectively.} \label{Fdiagram}\vspace*{.5cm}
\end{figure}

In principle, it would seem that the EFT at the LO described in Eq.~(\ref{LO-Lagran}) should be enough to compute to leading-order quantum and thermal corrections. 
However, this is not always the case.  When computing physical quantities close
to the phonon on-shell limit, and due to the fact that phonons remain always massless degrees of
freedom, one may detect sensitivity to the NLO dispersion law, even
when computing the leading-order corrections.

 In the following sections we study the main corrections
to the phonon dispersion law as derived from the EFT here described. We then see  an explicit
example where the problem discussed previously appears.

\section{One-loop phonon self-energy with the LO Lagrangian}
\label{1loopSec}

In this section we compute the phonon self-energy at
one-loop, as derived from the  LO EFT described in the previous section.
 We study separately
the  nonthermal and the thermal corrections, which we denote as
 $\Pi_{T=0}(P)$ and $\Pi_T (P)$,  respectively. As we will
see, in the low-momentum limit, the first are very suppressed.

The computation is done  using the imaginary time formalism (IFT) \cite{LeBellac}.
That is, one first performs a rotation to Euclidean time.
We  denote with
capital letters the four Euclidean momentum. Feynman rules for the phonon propagator and for vertices are
straightforwardly derived from the effective Lagrangian equation [Eq.(\ref{comlag})].
We use dimensional regularization to deal
with the ultraviolet (uv) divergences of the $T=0$ part of
the diagrams, so that $d= 4 - \epsilon$.

The phonon propagator in the ITF  with momentum
$Q^\mu=	  ( i \omega_n , {\bf q})= (q_0,{\bf q}) $ reads
 \be
\label{phonon-propag}
S(Q) \equiv \frac{1}{\omega_n^2 + E^2_q}  \ ,
\ee
 where $\omega_n = 2 \pi nT$, with $n \in \cal Z$,
is a bosonic Matsubara frequency. At the LO, one has
 $E_q = c_s q$, where $q= |{\bf q}|$.

There are three different diagrams that contribute to
the one-loop self-energy.
For external Euclidean momentum $P^\mu =  (i \omega, {\bf p})= (p_0,{\bf p})$,
these are the bubble, the tadpole, and lollypop diagrams
(see Fig. \ref{Fdiagram}).

The bubble and the tadpole diagrams are expressed as
 \be
\label{bubble}
 \Pi^{(b)} (P)	 =    -18 g^2  T \sum_{n=-\infty}^{\infty}
 \int
\frac{d^{d-1} q}{(2 \pi)^{d-1}}   \, \Big( F_1(P,Q) S(Q) S(P-Q)
\Big) \ ,
\ee
and 
\be
\label{tadpole}
\Pi^{(t)} (P)	=  -  \lambda
  T \sum_{n=-\infty}^{ \infty}
 \int
\frac{d^{d-1} q}{(2 \pi)^{d-1}} 
 \, F_2(P,Q) S(Q) \ ,
 \ee
respectively, where the vertex functions are
\begin{eqnarray}
\label{cub-vertex}
 F_1(P,Q) & = & [(p_0-q_0)p_0q_0-\eta_g\left((p_0-q_0){\bf p}\cdot{\bf q}+p_0({\bf p}\cdot{\bf q}-q^2)+q_0(p^2-{\bf p}\cdot{\bf q}) \right)]^2 \ ,\\
F_2(P,Q) & = & 12p_0^2q_0^2-2\eta_{\lambda,1}(4p_0q_0 \,{\bf p}\cdot {\bf q}+p_0^2q^2+q_0^2p^2)+4\eta_{\lambda,2}(2({\bf p}\cdot{\bf q})^2+p^2q^2) \ .
\end{eqnarray}

The lollypop diagram is expressed as 
 \be
\label{lolly}
 \Pi^{(l)} (P)	 =    18 g^2  T \sum_{n_q=-\infty}^{ \infty}
 \int
\frac{d^{d-1} q}{(2 \pi)^{d-1}}   \, \Big( H(P,Q) S(Q) 
\delta^{(4)}(Q) \Big)  T \sum_{n_k=-\infty}^{ \infty}
 \int
\frac{d^{d-1} k}{(2 \pi)^{d-1}}   \, \Big( H(K,Q) S(K) 
\Big) \ ,
\ee
where the vertex function is
\be
H(P,Q) = q_0 p_0^2 - \eta_g \left(q_0  p^2 + 2 q_0  {\bf p}\cdot{\bf q}   \right) º ,
\ee
and the four-dimensional delta function is defined as
\be
\delta^{(4)}(Q) = \frac{1}{T} \delta_{\omega_{n_q}} (2 \pi)^3 \delta^{(3)}({\bf q}) \ ,
\ee
where $\delta_{\omega_{n_q}}$ is a delta of Kroneker.

\subsection{One-loop self energy at $T=0$.}
\label{T0-oneloop}

By dimensional analysis, it is easy to check that  the one-loop corrections to the self-energy
are proportional to $P^6 \log{P}$ at $T=0$, and thus, they are very suppressed, particularly
with respect to the thermal corrections we show later on. 
However, we still discuss how these small corrections might affect
the phonon dispersion law at $T=0$.
 
In dimensional regularization the tadpole and lollypop diagrams are strictly zero. Thus,   at $T=0$ we only
need to consider the bubble diagram:
\begin{equation}
\label{bubbleT=0}
\Pi^{(b)}_{T=0}(P)= - 18g^2\int\frac{\,d^dQ}{(2\pi)^d}\frac{F_1(P,Q)}{(q_0^2+c_s^2q^2)((\omega-q_0)^2+c_s^2({\bf p}-{\bf q})^2)} \ .
\end{equation}
Is it convenient to make use of Feynman parameters. Thus, we perform the change of variables $R = Q- P x$ and rescale this last variable, doing  $r_0 = c_s z_0$ and ${\bf r} = {\bf z}$, and transform Eq.~(\ref{bubbleT=0}) into
\begin{equation}
\label{bubbleT=0-2}
\Pi^{(b)}_{T=0}(P)= -\frac{18g^2}{c_s^3}\int\frac{\,d^dZ}{(2\pi)^d}\int_0^1 dx\, \frac{\,F_1(P,Z)}{[z_0^2+z^2+\frac{\omega^2+c_s^2p^2}{c_s^2}x(1-x)]^2}  \ . 
\end{equation}

We perform first the integral in the $Z$ variable, using the well-known formula \cite{Peskin}
\begin{equation}
\label{tabul-integral}
\int\frac{\,d^d Z}{(2\pi)^d}\frac{1}{(Z^2+\Lambda)^2}=\frac{1}{(4\pi)^{d/2}}\frac{\Gamma(2-\frac{d}{2})}{\Gamma(2)}\left(\frac{1}{\Lambda}\right)^{2-\frac{d}{2}} 
\end{equation}
and similar ones with different powers of $Z_\mu$ in the numerator.
Here, as we are only interested in getting the lowest corrections to the phonon dispersion law,  we approximate the
numerator of Eq.~(\ref{bubbleT=0-2}), taking only the value for $Z=0$, thus
\begin{equation}
 F_1(P,0) = - x^2 (1-x)^2 \omega^2 (\omega^2 + 3 \eta_g p^2)^2 \ .
\end{equation}
The point is that if one considers finite powers of $Z$ in the numerator of Eq.~(\ref{bubbleT=0-2})
and performs the $Z$ integral, one gets more powers of $\Lambda \sim (\omega^2 + c_s^2 p^2)$ than
those that appear in Eq.~(\ref{tabul-integral}). We content ourselves with obtaining the lowest
 corrections to the dispersion law. In this limit, we also approximate $F_1(P,0)$ in a $(\omega^2 + c_s^2 p^2)$ expansion. Using the $\bar{MS}$ substraction scheme, and
after analytical continuation $\omega=-ip_0+0^+$,  we get
\begin{equation}
\label{T0oneloop}
\Pi^{(b)}_{T=0}(p_0,{\bf p}) \approx \frac{c_s p^6}{240\pi^2\rho_0 } 
\left(\frac{\rho_0}{c_s}\frac{\partial c_s}{\partial\rho_0}+1\right)^2 
  \left[\log\left(-\frac{p_0^2-c_s^2p^2}{c_s^2M_r^2}-ip_00^+\right)-\frac{47}{30}\right]
+ {\cal O} (p_0^2 -c_s^2 p^2) \ ,
\end{equation}
where we used the relations Eqs.~(\ref{rel-2}) and (\ref{rel-1})   of Appendix~\ref{set-relations}.

\subsection{Thermal corrections to the self-energy}
\label{T-oneloopcor}

For the computation of the thermal component of the self-energy one can take $d=4$, as it is
uv finite. As we are only interested in obtaining the leading corrections to the
dispersion law at low momentum, we only compute the self-energy in the limit $p_0, {\bf p} \rightarrow 0$.
In particular, we assume that these two scales are smaller that the scale set by the temperature.

In the low-external-momentum limit several approximations can be performed. Due to the
presence of the Bose-Einstein distribution function, it is clear that the leading contribution
to the integral comes from the region where $q \sim T$. In such a case, it is legitimate to
approximate
\begin{equation}
\label{approx-vertex}
 F_1(P,Q) \approx \left( -p_0 (q_0^2 - \eta_g q^2) - 2 \eta_g q_0 {\bf p}\cdot{\bf q} \right)^2 
\end{equation}
in the numerator of the bubble diagram, and also
\begin{equation}
\label{E1E2}
E_1 - E_2 \approx c_s \, p \cos{\theta} \ , \qquad f_1 - f_2 \approx c_s \, p  \cos{\theta} \,
\frac{d n_B(E_q)}{d E_q} \ .
\end{equation}

We present in Appendix~\ref{App-LOthermal} the explicit details of the computation of the
three Feynman diagrams, and here we only display the final result, valid in the  $p_0, {\bf p} \rightarrow 0$
limit: 
\begin{eqnarray}
\label{final-total}
\Pi_T(p_0, {\bf p}) & \approx &  -\dfrac{2 \pi^2 T^4}{15 c_s^5} \frac{ c_s^2p^2 }{\rho_0}
\Bigg [ \frac{1}{2} + \frac{1}{2}\frac{\rho_0}{c_s} \frac{\partial c_s}{\partial \rho_0}
+ \frac{p_0^2}{c_s^2 p^2} \left( -\frac{\rho_0}{c_s} \frac{\partial c_s}{\partial \rho_0} +
\frac{2\rho_0^2}{c_s^2} \left( \frac{\partial c_s}{\partial \rho_0}\right)^2 -\frac{\rho_0^2}{4c_s}\frac{\partial^2c_s}{\partial\rho_0^2}
 \right.
 \nonumber
 \\
 &+ &
 \left. \left. \left(\frac{\rho_0}{c_s}\frac{\partial c_s}{\partial\rho_0}+1\right)^2
  \right)
  +  \frac 12\left(\frac{\rho_0}{c_s}\frac{\partial c_s}{\partial\rho_0}+1\right)^2
   \left(\frac{p_0}{c_sp}\right)^3
   \log{\dfrac{p_0 - c_s p}{p_0+ c_s p}}
    \right] \ .
    \end{eqnarray}

As clearly seen from Eq.~(\ref{T0oneloop}), for low external momentum,  assuming
that $p \ll T$, the $T=0$ self-energy is very suppressed with respect to the thermal part
of the self-energy.

\section{Leading corrections to the self-energy coming from the NLO Lagrangian}
\label{1loopNLO}

In order to study both the thermal and the higher momentum corrections to the
linear phonon dispersion law,  knowledge of the phonon self-energy is necessary.
Naively, one could think that computing the one-loop self-energy with the LO
Lagrangian should be enough to get the leading corrections to Eq.~(\ref{lin-dlaw}). However, given the explicit form 
 of both $\Pi_T(P)$ [see Eq.~(\ref{final-total})] and 
 $\Pi_{T=0}(P)$ [see Eq.~(\ref{T0oneloop})], we see that when studied close to the on-shell limit,
these might be logarithmically sensitive to the NLO dispersion law as well.  If the corrections provided
by the one-loop LO physics to the phonon dispersion law are smaller than those associated with the value
of the parameter $\gamma$, see Eq.~(\ref{displaw-NLO}), then one cannot neglect the 
NLO Lagrangian in the first corrections to the dispersion law. We will assume that this is the case,
as the thermal corrections are governed by the ratio $(T^4/\rho_0 c_s^5)$, and the $T= 0$ corrections are governed
by the ratio $(p^4/\rho_0)$, and these two ratios are very small in the low-temperature regime of the superfluid.
In this situation, we must start our computation with the Lagrangian described 
by  $\mathcal{L}_{\rm LO} +\mathcal{L}_{\rm NLO}$.

In this section we compute the leading thermal and nonthermal corrections of the theory
described by  $\mathcal{L}_{\rm LO} +\mathcal{L}_{\rm NLO}$. Several new one-loop diagrams
appear, but most of them 
will produce  higher $p$ or $T$ corrections than those computed
in the previous section. Those are the diagrams that arise from higher-momentum corrections
to the different vertices of the LO physics, or new vertices with higher-momentum dependence.
We do not compute these  higher-order corrections here, but simply focus on getting the
corrections that contribute at the same order as those computed with $\mathcal{L}_{\rm LO}$.

It is very easy to spot the one-loop contributions that give the leading one-loop corrections.
Consider the bubble diagram, with vertices arising
only from $\mathcal{L}_{\rm LO}$, but with the propagator computed from 
$\mathcal{L}_{\rm LO} +\mathcal{L}_{\rm NLO}$. The phonon propagator takes the same
form as in Eq.~(\ref{phonon-propag}), but now the energy is given by Eq.~(\ref{displaw-NLO}).

Let us first consider the thermal corrections to the self-energy. If we only consider
the low-external-momentum limit, the analysis turns out to be very simular to the one
we carried out in the preceding section. Carrying out the same steps and approximations,
one only has to replace  Eq.~(\ref{E1E2}) by
\begin{equation}
 E_1 - E_2 \approx c_s p \cos\theta \left(1 - 3 \gamma q^2 \right)  \ .
\end{equation}

After an analysis very similar to the one we carried out before
(see Appendix~\ref{App-LOthermal} for details), we reach the following
expression when the  phonon is nearly on-shell,  $p_0 \approx c_sp(1-\gamma p^2)$,
\begin{eqnarray}
\label{final-total-NLO}
\Pi_T(p_0, {\bf p})\Big |_{p_0 \approx c_sp(1-\gamma p^2)} & \approx &  -\dfrac{2 \pi^2 T^4}{15 c_s^5} \frac{ c_s^2p^2 }{\rho_0}
\Bigg [ \frac{1}{2} + \frac{1}{2}\frac{\rho_0}{c_s} \frac{\partial c_s}{\partial \rho_0}
+ \frac{p_0^2}{c_s^2 p^2} \left( -\frac{\rho_0}{c_s} \frac{\partial c_s}{\partial \rho_0} +
\frac{2\rho_0^2}{c_s^2} \left( \frac{\partial c_s}{\partial \rho_0}\right)^2 -\frac{\rho_0^2}{4c_s}\frac{\partial^2c_s}{\partial\rho_0^2}
\right.
\nonumber
\\
&+ &
  \left. \left. \left(\frac{\rho_0}{c_s}\frac{\partial c_s}{\partial\rho_0}+1\right)^2
     \right)
 +  \frac 12\left(\frac{\rho_0}{c_s}\frac{\partial c_s}{\partial\rho_0}+1\right)^2
     \left(\frac{p_0}{c_sp}\right)^3
        \log{\dfrac{27\gamma T^2+i0^+}{c_s^2}}
	    \right] \ .
\end{eqnarray}

A similar thing happens at $T=0$. The bubble diagram is modified when considering the
corrected phonon propagator. In particular, it now reads
\begin{equation}
\Pi^{(b)NLO}_{T=0}(P)= -\frac{18g^2}{c_s^3}\int\frac{\,d^dZ}{(2\pi)^d}\int_0^1\,dx\frac{F_1(P,Z)}{[z_0^2+z^2+\frac{\omega^2+c_s^2p^2}{c_s^2}x(1-x)+\gamma F_3(P,Z,x)]^2} \ ,
\end{equation}
where $F_3(P,Z,x)=-2[({\bf z}+x{\bf p})^4(1-x)+({\bf z}-(1-x){\bf p})^4x]$.
This expression can be simplified by the following argument: the term proportional to $\gamma$ in the denominator will only be of the same order of magnitude as the rest of the denominator for very small $Z^2$, but if this happens $P\gg Z$ and this means that the approximation $F_3(P,Z,x)\sim F_3(P,0,x)$ can be made. If one considers also that the external particle is
nearly on-shell,$\omega^2+c_s^2p^2 \approx 2c_s^2\gamma p^4$, then
\begin{equation}
\Pi^{(b)NLO}_{T=0}(P) \approx  - \frac{18g^2}{c_s^3}\int\frac{\,d^dZ}{(2\pi)^d}\int_0^1\,dx\frac{F_1(P,0)}{[z_0^2+z^2+6\gamma p^4x^2(1-x)^2]^2} \ .
\end{equation}
In the $\bar{MS}$ substraction scheme, and
after analytical continuation to Minkowski space, we get
\begin{equation}
\label{finalT0}
\Pi^{(b)NLO}_{T=0}(p_0, {\bf p})\Big |_{p_0 \approx c_sp(1-\gamma p^2)} \approx 
\frac{c_sp^6}{240\pi^2\rho_0}\left(\frac{\rho_0}{c_s}\frac{\partial c_s}{\partial\rho_0}+1\right)^2\left[\log\left(\frac{6\gamma p^4}{M_r^2}-ip_00^+\right)-\frac{47}{15}\right]  \ .
\end{equation}

\section{The phonon dispersion relation}
\label{phdisp-sec}

The phonon dispersion law is obtained by studying the poles of the resumed propagator,
that is, after solving
\begin{equation}
\label{dispersion-law}
 p_0^2 - p^2 c_s^2 - \Pi(p_0,{\bf p}) = 0 \ ,
\end{equation}
where $\Pi$ is the one-loop self-energy. Because the $T=0$ contributions to this
quantity are very suppressed, we first only consider the temperature-dependent part.

We divide Eq.~(\ref{dispersion-law})
by $p^2 c_s^2$ and realize that the real part of the equation
only leads to thermal corrections to the phonon velocity. After writing
$p_0 \approx \left( c_s + \delta v(T) \right) p-i\alpha(p,T)$, we reach
\begin{equation}
\label{velfonon}
\delta v(T) \approx    -\frac{\pi^2}{15 \rho_0} \left( \frac{T}{c_s}\right)^4
\left[ - \frac 14 \frac{\rho_0^2}{c_s}\frac{\partial^2 c_s}{\partial\rho_0^2}  + \frac{1}{2}
- \frac 12 \frac{\rho_0}{c_s} \frac{\partial c_s}{\partial\rho_0} +
 \frac{2\rho_0^2}{c_s^2} \left( \frac{\partial c_s}{\partial\rho_0}\right)^2
 + \left( 1 + \frac{\rho_0}{c_s} \frac{\partial c_s}{\partial\rho_0}\right)^2 \left(1 +
 \frac 12 \log{\frac{27 |\gamma| T^2}{c_s^2}} \right)
  \right] 
  \end{equation}
and
\begin{equation}
\label{damping}
\alpha(p,T)= \Theta(-\gamma) \frac{T^4p\pi^3}{30\rho_0c_s^4}\left(1+\frac{\rho_0}{c_s}\frac{\partial c_s}{\partial\rho_0}\right)^2 \ .
\end{equation}

The leading thermal corrections we have obtained,
that is, those that go as $\sim T^4 \ln{T}$ are the same as those obtained from
the kinetic theory approach of Andreev and Khalatnikov \cite{Andreev}, which we review
in Sec.~\ref{kinetic+hydro}.

For completeness, we also discuss the modifications to the dispersion law at $T=0$ that arise
from the one-loop physics. If we assume $p_0 \approx  c_s  p \left(1 - \gamma p^2 \right) + \delta E_p-i\alpha(p)$, then we obtain
\begin{equation}
\delta E_p=\frac{ c_s p^5}{480\pi^2\rho_0}\left(\frac{\rho_0}{c_s}\frac{\partial c_s}{\partial\rho_0}+1\right)^2\left[\log\left(\frac{2 |\gamma| p^4}{M_r^2}\right)-\frac{47}{15}\right] 
\end{equation}
and
\begin{equation}
\alpha(p)= \Theta(-\gamma) \frac{p^5}{480\pi \rho_0}\left(\frac{\rho_0}{c_s}\frac{\partial c_s}{\partial\rho_0}+1\right)^2 \ .
\end{equation}

As for the damping terms, we note that if we work with the LO phonon Lagrangian,
we obtain the same damping term, but without the step function of $-\gamma$.
For the case of a weakly interacting Bose gas, where the following property
is fulfilled
$
\frac{\rho_0}{c_s}\frac{\partial c_s}{\partial\rho_0} =\frac{1}{2}
$ (see, for example Ref.~\cite{Andersen:2003qj}),
we then get the results for the damping term first obtained in Refs.~\cite{Hohenberg,Beliaev,Popov}.
However, when working to the NLO, with a dispersion law of the form expressed in
Eq.~(\ref{displaw-NLO}), the process described by $\alpha$, which corresponds
to one phonon decaying into two, is only kinematically allowed for negative
values of $\gamma$.

The thermal corrections to order $T^4 \ln{T}$  of the phonon velocity of a weakly interacting Bose gas
have been computed in Ref.~\cite{Giorgini} [see  Eq.~(54)]. We note that we agree with the
value of the coefficient in front of the logarithm. However, Giorgini \cite{Giorgini} computed only at the LO
and thus does not show any dependence of the result on $\gamma$, while
our computation proceeds at the NLO and shows the sensitivity of these logarithmic corrections to
the nonlinear corrections to the phonon dispersion law, which cannot be neglected.

\section{Thermal corrections to the speed of sound in a cold superfluid from kinetic theory}
\label{kinetic+hydro}


The thermal corrections to the speed of	 sound in a $^4$He  superfluid	at low $T$ were first computed by Andreev and
Khalatnikov \cite{Andreev} in a collisionless regime. For completeness, we review such a computation here
(see also Ref.~\cite{IntroSupe}) and extend it to get also the subleading thermal corrections.

We consider  Landau's two fluid  equations for the superfluid \cite{landaufluids,IntroSupe}
to study the evaluation of the dynamics of sound waves.
For our purposes, it is enough to concentrate only on
the continuity equation and the equation for the superfluid velocity ${\bf v}_si$:
\begin{eqnarray}
 \label{continuity}
\partial_t \rho + {\rm div} {\bf j}  & = &  0  \ ,
\\
\label{iflow}
\partial_t {\bf v_s} + \nabla \left(\mu + \frac{{\bf v}_s^2}{2}\right) & = & 0\,,
\end{eqnarray}
where $\rho$ is the mass density, ${\bf j}$ is the current, and $\mu$ is the chemical potential.

At zero temperature only the pure superfluid component exists. At
 finite temperature, a normal fluid component has to be considered as well,
which takes into account the contribution to the hydrodynamics of the different
 quasiparticles. However, at very low temperatures only the massless quasiparticles of the 
superfluid, the phonons,
are  thermally excited and contribute
to the thermodynamical and hydrodynamical properties of the system.

In the very-low-$T$ regime the mean free path of the superfluid phonons is very
large \cite{IntroSupe}, and one can consider that they are in a collisionless regime or, in other words, that they are not in a hydrodynamical regime. In this case the
thermal phonons	can be viewed as forming a bosonic gas,  and thus they can be described with a Boltzmann equation.

In the collisionless regime
the phonon distribution function $n$  obeys the kinetic	 equation \cite{IntroSupe}
\begin{equation}
\label{Boltzman}
\frac{\partial n}{\partial t} + \frac{\partial n}{\partial {\bf r}} \cdot \frac{\partial H}{\partial {\bf p}}
- \frac{\partial n}{\partial {\bf p}} \cdot \frac{\partial H}{\partial {\bf r}} = 0 \,,
\end{equation}
where $H= E_p + {\bf p} \cdot {\bf v}_s$ is the phonon Hamiltonian. The energy of the phonon
can be taken as in Eq.~(\ref{displaw-NLO})
if we move from the strict low-momentum limit. The phonon transport equation for the
weakly interacting Bose gas has been derived in Ref.~\cite{Kirkpatrick} (see also Ref.~\cite{Griffin}).

The aforementioned transport equation has been explicitly derived for a weakly interacting Bose
gas, starting from the microscopic description of the gas \cite{Kirkpatrick}. We are
not aware of a similar derivation in the literature for other superfluid systems.

The phonons give contributions to the current and chemical potential of the system,
which are expressed as
\begin{equation}
\label{current}
{\bf j} =  \rho {\bf v}_s + \tint {\bf q}\, n \ , \qquad
\mu  =  \mu_0 +   \tint \frac{\partial E_q}{\partial \rho} \, n \ ,
\end{equation}
where $\mu_0$ is the chemical potential at $T=0$. 

One then studies deviations from the equilibrium solutions, proportional to $e^{-i \omega t+i {\bf k} \cdot {\bf r}}$, so that
\begin{equation}
n= \bar n+ \delta n	 \ , \qquad
\rho= \bar \rho + \delta \rho \ ,
\end{equation}
where the quantities with the bar  refer to the equilibrium values.
 For the unperturbed value of the phonon distribution function we take
the Bose-Einstein distribution function, $\bar n = n_{B}(E_p)$.

We also assume that we work in the frame where
$\bar {\bf v}_{s} =0$, but there is also a fluctuation in the superfluid velocity, $\delta {\bf v}_s$.
In Ref.~\cite{IntroSupe}, it is mentioned that one should also consider the functional dependence
of the phonon energy on the density of the excitations, with a term of the form
$\delta E_p = \int_{\bf p'} f({\bf p},{\bf p'})  \delta n(p')$, where $f$ is a function to be determined. This term was neglected by Andreev and Khalatnikov \cite{Andreev}
under the assumption that it gives very subleading corrections to the final result. 
For the time being, we make the same assumption. However, in order to check the
accuracy of this assumption, one might need to evaluate the form of the function $f$, starting
from the  microscopic description of the superfluid.

After linearizing the kinetic equation (\ref{Boltzman}), one finds the value of the fluctuation
in terms of $\delta \rho$ and $\delta {\bf v}_s$:
\begin{equation}
\delta n=-\frac{d n_B}{d E_p } \frac{{\bf k}\cdot {\bf v}_{\rm ph}}{\omega - {\bf k}\cdot {\bf v}_{\rm ph}  }\left(\frac{\partial E_p }{\partial\rho} \delta \rho + {\bf p}\cdot \delta {\bf v}_s
\right) \ ,
\end{equation}
where the phonon velocity is given by ${\bf v}_{\rm ph} = \frac{\partial E_p}{\partial {\bf p}}$.
Consequently, one can compute the fluctuations in both the current and the chemical potential due to
these phonon fluctuations using Eqs.~(\ref{current}).  After linearizing also the
hydrodynamical equations, Eqs.~(\ref{continuity}) and (\ref{iflow}), one gets a homogeneous system of linear equations for $\delta \rho$ and $\delta {\bf v}_s$. Keeping only the leading and subleading thermal corrections, and assuming $v_{\rm ph} \approx c_s$, one finds
\begin{eqnarray}
\label{disp-law-sound-2}
\nonumber
-\frac{\omega^2}{k^2} + c_s^2  &= & \int^\infty_0 \frac{p^4 dp }{2\pi^2}\frac{d n_B}{d E_p} \left\lbrace 	 \left(c_s\frac{\bar \rho}{4}\frac{\partial^2 c_s}{\partial\bar \rho^2}
- \bar \rho \left( \frac{\partial c_s}{\partial\bar \rho}\right)^2
 -  \frac{c^2_s}{\bar \rho}  \left( \frac 13 +
\frac{\omega^2}{k^2 v^2_{\rm ph}} \right)
- 2 \frac{\partial c_s}{\partial\bar \rho} \frac{\omega^2}{k^2 v_{\rm ph}} \right)
\right.
 \\
&- &
\left.
\frac 12
\left(
- \bar \rho \left( \frac{\partial c_s}{\partial\bar \rho}\right)^2   \frac{\omega}{k v_{\rm ph}}  -
\frac{c^2_s}{\bar \rho}   \frac{\omega^3}{k^3 v^3_{\rm ph}} - 2 \frac{\partial c_s}{\partial\bar \rho}   \left( \frac{\omega^3}{k^3 v^2_{\rm ph}} \right)  \right)
\int^1_{-1} \frac{ d \cos{\theta}}{ \frac{\omega}{k v_{\rm ph}} - \cos{\theta} + i0^+} 
\right\rbrace 
\ ,
\end{eqnarray}
and retarded boundary conditions have been assumed, with the $+i0$ prescription.

Equation (\ref{disp-law-sound-2}) might be interpreted as the dispersion law for the  hydrodynamical wave.
Its solution contains both real and imaginary parts, and the first is seen to correct only the
value of the speed of sound. Thus, if we define
$\omega \approx (c_s + \delta c_s(T)) k-i\alpha(T,k)$, 
where $\delta c_s$ is the correction to the sound velocity due to the thermal fluctuations
 and $\alpha$ is the sound decay coefficient, or attenuation factor, one then gets
\begin{equation}
\label{velsonido}
\delta c_s(T) \approx  - \frac 12 \frac{c_s}{\bar \rho} \int^\infty_0 \frac{p^4 dp }{2\pi^2}\frac{d n_B}{d E_p}
\left[  \frac 14 \frac{\bar \rho^2}{c_s}\frac{\partial^2 c_s}{\partial\bar \rho^2}  - \frac 13
- \left( 1 + \frac{\bar \rho}{c_s} \frac{\partial c_s}{\partial\bar \rho}\right)^2 \left(1 + \frac 12 \log{\frac{\omega - k v_{\rm ph}}{\omega + kv_{\rm ph} }} \right) 
 \right] \ .
\end{equation}

Andreev and Khalatnikov \cite{Andreev} solved the dispersion relation, Eq.~(\ref{velsonido}), keeping only the leading terms
in $T$. They noticed that the solution was sensitive 
to the form of the phonon dispersion law. Eq.~(\ref{displaw-NLO}), and the parameter $\gamma$.
The equation can also be solved keeping subleading terms in $T$.  One finds
\begin{equation}
 \delta c_s(T) \approx    \frac{\pi^2 }{30 \bar \rho}  \left(  \frac{T}{c_s} \right)^4
 \left(1+\frac{\bar \rho}{c_s}\frac{\partial c_s}{\partial\bar \rho}\right)^2
\log{ \frac{c_s^2}{27 |\gamma| T^2}} -  \frac{ \pi^2 }{15 \bar \rho}  \left( 
 \frac{T}{c_s} \right)^4 
\left(-\frac 14 \frac{\bar \rho^2}{c_s}\frac{\partial^2 c_s}{\partial\bar \rho^2}
+ \frac 13 + \left(1+\frac{\bar \rho}{c_s}\frac{\partial c_s}{\partial\bar \rho}\right)^2 
\right) \ .
\end{equation}

We also find a nonvanishing attenuation factor if the parameter $\gamma$ is negative:
\begin{equation}
\label{damping1}
\alpha(T,k)= \Theta(-\gamma) \frac{ \pi^3  k }{30 \bar \rho} 
\dfrac{T^4}{c_s^4}\left(1+\frac{\bar \rho}{c_s}\frac{\partial c_s}{\partial\bar \rho}\right)^2  \ ,
\end{equation} 
 where $\Theta$ is the step function. Andreev and Khalatnikov \cite{Andreev} did not consider such a term,
as they assumed that $\gamma$ was positive for $^4$He.

We agree with Andreev and Khalatnikov \cite{Andreev} in the correction of order $T^4 \ln{T}$ to the speed of
sound, except for
a factor 2 in the argument of the logarithm, as already pointed out in Ref.~\cite{Singh},
which is probably due to the approximation that these authors used to solve the integral of
Eq.~(\ref{velsonido}).
Here instead the corresponding integral was computed exactly.

Because  at low $T$ one can approximate $\bar \rho \approx \rho_0$, we see that the corrections of
order $T^4 \ln{T}$ are the same as those obtained with the EFT for the phonon velocity [see Eq.~(\ref{velfonon})].
As for the corrections of order $T^4$,  there is a clear discrepancy between
the two results.  However the discrepancy involves the same kind of thermodynamical derivatives, an
indication that the two approaches must be very similar.
  We believe that this discrepancy might require one to have a rigorous
derivation of the phonon kinetic theory approach for every superfluid system, so as to derive all possible
sources of the subleading thermal corrections to different physical quantities.

\section{Applications to the cold unitary Fermi gas}
\label{coldFgas-sec}

In the unitarity limit, the thermodynamic properties of the cold Fermi gas can be determined
up to some dimensionless constants \cite{Ho:2004zza}. At  zero temperature, and
due to the absence of any internal scale, dimensional analysis fixes the form of the
pressure as being proportional to that of a  free system,
\be
P = c_0 m^{3/2} \mu_0^{5/2} \,,
\ee
where $c_0$ is a dimensionless and universal constant.
This parameter can  be expressed as
\be
c_0 = \frac{2^{5/2}}{15 \pi^2 \xi^{3/2}}\,,
\ee
where  $\xi$ is the universal constant  that fixes the relation between the chemical potential and the Fermi energy $\mu_0 = \xi E_F$. Experiments with cold trapped fermionic atoms~\cite{Bartenstein:2004zza} find $\xi \sim 0.32 -0.44$, a result that is in agreement with 
Quantum Monte Carlo calculations at vanishing temperature~\cite{Chang:2004zz,Astrakharchik:2004zz}. 

At sufficiently low temperatures, the system is a superfluid phase. The temperature for the superfluid 
phase transition has been determined also with Monte Carlo numerical simulations, and then one finds
 $T_c \sim 0.23 E_F = \frac{0.23}{\xi} \mu_0$ \cite{Bulgac:2005pj}.

At zero temperature one can  easily deduce the density
from the value of the pressure, as well
as the value of the speed of sound, which is $c_s^2 = 2 \mu_0/3 m$.
In this case it is easy to check 
\begin{equation}
\frac{\rho_0}{c_s}\frac{\partial c_s}{\partial\rho_0}=\frac{1}{3} \ , \qquad 
\frac{\rho^2_0}{c_s}\frac{\partial^2 c_s}{\partial\rho_0^2}= -\frac{2}{9}
\end{equation}
and that the combination
\begin{equation}
c_s^5\rho_0=\frac{16\mu_0^4}{27\sqrt{3}\pi^2\xi^{3/2}} 
\end{equation}
only depends on $\mu_0$ and on $\xi$, but not on the value of the mass of the specific fermion species
one is considering.

The values of $\xi$ and $\gamma$ can be extracted from a fit to numerical Monte Carlo simulations~\cite{salasnich}.
Following the notation of Ref.~\cite{salasnich}, we can parametrize
\begin{equation}
\frac{\gamma}{c_s^2}=-\frac{\lambda}{8m^2c_s^4}=-\frac{9\lambda}{32\mu_0^2} \ ,
\end{equation}
and $\lambda$ is also an adimensional universal constant. The fit to numerical simulations that was performed in Ref.~\cite{salasnich}  gives the
result  $\xi=0.455$ and $\lambda=0.13$.

We can express the relative thermal correction of the phonon velocity as 
\begin{equation}
\label{rel-phvelo}
\frac{\delta v(T)}{c_s}=-\frac{\pi^4\sqrt{3}\xi^{3/2}}{160}\left(\frac{T}{\mu_0}\right)^4
\left\{43+16\log\left(\frac{243\lambda}{32}\frac{T^2}{\mu_0^2}\right)\right\} \ ,
\end{equation}
which depends only on the  ratio $\frac{T}{\mu_0}$ and the constants $\xi$ and $\lambda$.
It is important to stress the universal character of this thermal correction for all cold Fermi
gases at unitarity.

\begin{figure}[!th]
\includegraphics[width=2.5in,angle=-0]{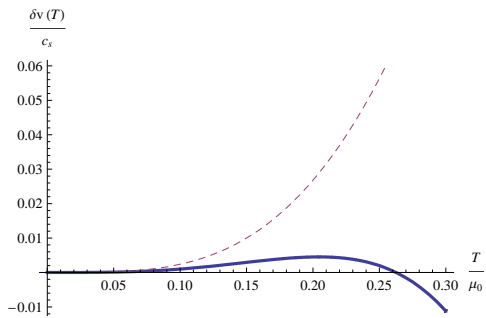}
\caption{Relative thermal correction to the phonon velocity for a cold Fermi gas in 
the unitarity limit. With the dashed line we plot the corrections of order $T^4 \ln{T}$, which agree with
that of the speed of sound  predicted with
kinetic theory. With the solid line, we plot the entire correction that includes the subleading terms of order $T^4$, given in Eq.~(\ref{rel-phvelo}). 
We use the values  $\xi=0.455$ and $\lambda=0.13$ obtained in Ref.~\cite{salasnich}.
} \label{figure2}
\end{figure}

In Fig.~\ref{figure2}, we plot Eq.~(\ref{rel-phvelo}), using the numerical the values of $\xi$ and $\lambda$ obtained in Ref.~\cite{salasnich}, up to temperatures below $T_c$, the critical temperature for the transition
to the normal phase.  As we see from the plot, the corrections of order $T^4$ compensante those
of order $T^4 \ln{T}$ and make the thermal correction to the phonon velocity really tiny
at still relatively low temperatures. At sufficiently low temperatures, one may expect that
these corrections are the same as those for the speed of sound, as the phonons are the only
thermal excited states. At  temperatures close to $T_c$ 
one may expect a discrepancy between these two quantities, as other quasiparticles (for example,
the fermion pairs) may contribute to the corrections to the speed of sound.

The sound attenuation factor is also given by
\begin{equation}
\label{damping-coldFermigas}
\alpha(p,T)=  \frac{8T^4p\pi^3}{135\rho_0c_s^4} \ .
\end{equation}

At larger temperatures, when phonons are in a hydrodynamical regime, the expression for
the sound attenuation is different, as it is expressed in terms of the different transport
coefficients of the system \cite{IntroSupe}. For cold Fermi gases the sound attenuation  
in the phonon hydrodynamical regime  has been recently computed in Ref.\cite{Braby:2010ec}.
It is not clear at which values of $T$ one could consider that the phonons are in a hydrodynamical
regime or not, as this would depend on the ratio between the phonon mean free path and the macroscopic volume
of the system.

\section{Conclusions}
\label{conclu}


In this article we have studied the phonon dispersion law in an arbitrary superfluid system as derived from the phonon EFT
in a collisionless or, equivalently, very-low-$T$ regime. We have concentrated our efforts on considering
how the speed of the phonon is modified at low temperature. Because the phonons are the quanta of the sound waves in the
superfluid, when looking into this modification in the low-momentum limit, such a thermal correction $\delta v(T)$
should be the same as the one corresponding to the speed of sound, $\delta c_s(T)$, as in this
regime the phonons are the only  excited states.

We have compared the results of $\delta v(T)$ with the old computation of $\delta c_s(T)$ due to Andreev and Khalatnikov \cite{Andreev},
which is based on studying the superfluid hydrodynamical equations, together with a kinetic transport equation for the
phonons. The last was deduced based on very general principles, assuming the {\it ad hoc} phonon Hamiltonian of Landau.
It has only been derived from the microscopic theory for the superfluid phase of a weakly interacting Bose gas.
For very low $T$, our results agree with those of Andreev and Khalatnikov \cite{Andreev} in the corrections of order $T^4 \ln{T}$.
However, when we push the kinetic theory approach to higher-order corrections in $T$, we find a different result
than that predicted with EFT techniques. We believe that the origin of this discrepancy might require a modification
in the standard kinetic theory approach. Possible corrections might be due to the change in the phonon spectrum in
the presence of fluctuations of the density of  the phonon excitations, a term already discussed in Ref.~\cite{IntroSupe}.
It is, however, not clear how to derive such corrections in the kinetic theory approach without studying the specific underlying
microscopic system under consideration. On the other hand, the EFT techniques we used are based on simple symmetry and power counting
arguments, which are universally valid for all superfluid systems.

Apart from demonstrating the efficiency of the EFT techniques,
our ultimate goal is considering a  well-defined framework
to study the  phonon contributions to
the thermodynamical and hydrodynamical properties of the superfluid, especially those which
are not weakly coupled, such as the Fermi liquid close to the unitarity limit.
Very recently, different transport coefficients for the superfluid regime of these systems
have been computed \cite{Rupak:2007vp,Escobedo:2009bh,Braby:2010ec}. The computations were
done using the kinetic theory described by Khalatnikov \cite{IntroSupe} and used here. While all computations were done to the
leading order in a temperature expansion, 
 it might be interesting to see whether modifications
to the kinetic approach are needed if one wants to study subleading corrections in $T$ of
the transport coefficients. Alternatively, one could compute the different phonon contributions to the
transport coefficients using the phonon EFT. However, as known for other systems  such as quark-gluon plasma \cite{Litim:2001db}, the consideration of collision terms in a field theory treatment
would require one to study the theory at the two-loop order.  
We leave all these important issues for future studies.

\appendix

\section{Some useful relations }
\label{set-relations}

In order to compare the results of the phonon dispersion law with  the results obtained for
the dynamics of the sound waves, it is convenient to express differently
the phonon coupling constants of the LO Lagrangian that appear in  Eqs.~(\ref{coup-constants}). While in
Eqs.~(\ref{coup-constants}) those are expressed in terms of ratios of derivatives of the
pressure with respect to the chemical
potential, it will turn out to be more convenient to express them as derivatives of the speed
of sound with respect to the density. 

Like in the main part of this article,  in this appendix we denote quantities evaluated at zero temperature with the subscript $0$.
From the definition of the speed of sound $c_s$, one can check that at zero temperature one
can write
\begin{equation}
 \frac 12 \frac{\partial c_s^2}{\partial \mu_0} = \frac{1}{2m} - \frac{1}{2m} \dfrac{\frac{\partial P}{\partial \mu_0}}{\left(\frac{\partial^2 P}{\partial \mu_0^2} \right)^2 } \dfrac{\partial^3 P}{\partial \mu_0^3} \ .
\end{equation}
Also from the same definition and that of the mass density $\rho_0$, one has
\begin{equation}
 \frac{\rho_0}{c_s^2} = m \frac{\partial \rho_0}{\partial \mu_0} = m^2 \frac{\partial^2 P}{\partial \mu_0^2}
\ .
\end{equation}
Using the previous two equations, one can immediately check that
\begin{equation}
\label{ant-rel-2}
  \frac{\rho_0}{c_s} \frac{\partial c_s}{\partial \rho_0} = \frac 12 - \frac{1}{2} \dfrac{\frac{\partial P}{\partial \mu_0}}{\left(\frac{\partial^2 P}{\partial \mu_0^2} \right)^2 } \dfrac{\partial^3 P}{\partial \mu_0^3} = \frac{1}{2} - \frac{c_s^2}{2 \eta_g} \ ,
\end{equation}
where $\eta_g$ is the phonon self-coupling defined in Eqs.~(\ref{coup-constants}).
From this expression we deduce the following relation
\begin{equation}
\label{rel-2}
 \dfrac{1}{4\eta_g^2}\left( 3 \eta_g -c_s^2\right)^2 = \left( \frac{\rho_0}{c_s} \dfrac{\partial c_s}{\partial \rho_0} + 1\right)^2  \ .
\end{equation}

From the definition of the coupling constants, see Eqs.~(\ref{coup-constants}), it is easy to infer
\begin{equation}
\label{rel-1}
g^2=\frac{c_s^2}{36\rho_0\eta_g^2} 
\end{equation}
and
\begin{equation}
\label{rel-3}
\lambda \eta_{\lambda,2}  = \dfrac{c_s^2}{8 \rho_0} \ .
\end{equation}

Also from Eqs.~(\ref{coup-constants}) and (\ref{ant-rel-2}) one can check that
\begin{equation}
\label{rel-4}
 \dfrac{c_s^2 \eta_{\lambda,1}}{\eta_{\lambda,2}} = 2 \frac{c_s^2}{\eta_g} = 2 - 4 \dfrac{\rho_0}{c_s}\dfrac{\partial c_s}{\partial \rho_0} \ .
\end{equation}

Another useful relation that is needed in the one-loop computations performed in this article is
\begin{equation}
\label{rel-5}
 \frac{c_s^4}{\eta_{\lambda,2}} = \frac{1}{3} \left( 1 - 8 \dfrac{\rho_0 }{c_s} \frac{\partial c_s}{\partial \rho_0}
+ 10 \dfrac{\rho_0^2 }{c_s^2} \left(\frac{\partial c_s}{\partial \rho_0}\right)^2 - 2 \dfrac{\rho_0^2 }{c_s} \frac{\partial^2 c_s}{\partial \rho_0^2} \right) \ ,
\end{equation}
which can be deduced after some algebra, from the definitions Eqs.~(\ref{coup-constants}), and after expressing
$\frac{\partial^4 P}{\partial \mu_0^4}$ in terms of derivatives of the speed of sound and of $\rho_0$  \ .

\section{Computation of the thermal one-loop self-energy}
\label{App-LOthermal}

We present in this appendix the explicit details of how to obtain the thermal component of the one-loop 
phonon self-energy in the limit of low external momentum.

We first consider the bubble diagram. After performing the sum of Matsubara frequencies
 one finds \cite{Manuel:2004iv}
\be \label{impart}
 \Pi^{(b)} (P)	 =   18 g^2 \sum_{s_1, s_2 = \pm}
 \int \frac{d^{d-1} q}{(2 \pi)^{d-1}} \,  F_1(P,Q) \Big |_{q_0 = s_1 E_1}
\Bigg(\frac{s_1 s_2}{4 E_1 E_2} \frac{1 + f(s_1 E_1) + f(s_2 E_2)}
{i \omega - s_1 E_1 -s_2 E_2} \Bigg) \ . \ee
 where $E_1 = E_q$ and
$E_2 =E_{|{\bf p - q}|}$, and
\be
 f (x) \equiv \frac{1}{e^{x/T} -1}
\ . \ee
We use the same notational conventions of Ref.~\cite{LeBellac}, so that the
 equilibrium Bose-Einstein
distribution function is $
n_B (E) = f(|E|)$. Further one can see that $f(-E) = - (1 + n_B (E))$.
We
evaluate separately the pieces of Eq.~(\ref{impart}) where $s_1 = - s_2$ and where $s_1 = s_2$,
taking into account only the thermal contributions to the integrals.
For the pieces where  $s_1 = - s_2$, we have

\begin{equation}
\label{bubble-1}
\Pi^{(b)}_{1,T}(P)	\approx -\frac{9 g^2}{2}   \int\frac{\,d^3q}{(2\pi)^3}\frac{d n_B}{d E_q}\frac{1}{E_q^2}\frac{c_s \cos\theta}{i \omega-c_sp\cos\theta}
\left[ F_1(P,Q)+F_1(P,-Q) \right] \Big |_{q_0 = E_q} \ ,
\end{equation}
where we have used the approximations given in Eq.~(\ref{E1E2}), and we also consider
that in the low-momentum external limit one can use Eq.~(\ref{approx-vertex}).
Now we proceed in exactly the same way with the pieces of the bubble diagram, Eq.~(\ref{impart}), when
 $s_1 = s_2$, considering only the thermal part of the diagram. In the limit of low
external momentum  we get
\begin{equation}
\label{bubble-2}
\Pi^{(b)}_{2,T}(P)	\approx - \frac{9 g^2} {2}  \int\frac{\,d^3q}{(2\pi)^3}\frac{ n_B(E_q)}{ E^3_q}
\left[ F_1(P,Q)+F_1(P,-Q) \right] \Big |_{q_0 = E_q} \ .
\end{equation}

At this point, we also perform the analytical continuation to Minkowski space with retarded boundary
conditions, $i\omega \rightarrow p_0 + i 0^+$, and thus we get
\begin{equation}
\Pi^{(b)}_{1,T} (p_0, {\bf p})
\approx - \frac{1}{8\eta_g^2 \rho_0}\int\,\frac{d^3q}{(2\pi)^3}\frac{d n_B}{d E} \frac{q^2 \cos\theta}{\frac{p_0}{c_sp}-\cos\theta+i 0^+}\left[2p_0^2(\eta_g^2-2\eta_g c_s^2+c_s^4)+8p_0c_sp(\eta_g^2-\eta_g c_s^2)\cos\theta +8\eta_g^2c_s^2p^2\cos^2\theta\right] \ ,
\end{equation}
where we used Eq.~(\ref{rel-1}).

It is then realized that this part of the diagram
has both real and imaginary parts, the last one being related
to the damping of the phonon.
Using  the relation Eq.~(\ref{rel-2})
of Appendix~\ref{set-relations}, it can be rewritten as
\begin{equation}
\label{final-bubble-1}
\Pi^{(b)}_{1,T} (p_0, {\bf p})\approx
 \frac{c_s^2p^2}{\rho_0}\int\, \frac{\,d^3q}{(2\pi)^3}\frac{d n_B}{\partial E_q} q^2
\left[\frac{1}{3}+\left(\frac{p_0}{c_sp}\right)^2 \left(\frac{\rho_0}{c_s}\frac{\partial c_s}{\partial\rho_0}+1\right)^2-\left(\frac{\rho_0}{c_s}\frac{\partial c_s}{\partial\rho_0}+1\right)^2\frac{\left(\frac{p_0}{c_sp}\right)^3}{\frac{p_0}{c_sp}-\cos\theta+i
0^+}\right] \ .
\end{equation}

As for Eq.~(\ref{bubble-2}), using the
the relation Eq.~(\ref{rel-1}), together with Eq.~(\ref{ant-rel-2}),
 it can be expressed  as
\begin{equation}
\label{final-bubble-2}
\Pi^{(b)}_{2,T} (p_0, {\bf p})\approx  \frac{c_s^2p^2}{\rho_0}
\int\,\frac{d^3q}{(2\pi)^3}\frac{d n_B}{d E} q^2
\left[\frac{1}{12}+\frac{1}{4}\left(\frac{p_0}{c_sp}\right)^2\frac{\rho_0^2}{c_s^2}\left(\frac{\partial c_s}{\partial\rho_0}\right)^2\right] \ .
\end{equation}

The frequency sum in the tadpole diagram Eq.~(\ref{tadpole}) can be done using standard techniques and gives
\begin{equation}
\Pi^{(t)}(P) = - \lambda\int\frac{\,d^{d-1}q}{(2\pi)^{d-1}}\sum_{s=\pm}F_2(P,Q) \Big |_{q_0=s E_q}\frac{s}{2 E_q}(1+ f(s E_q)) \ .
\end{equation}

After analytical continuation to Minkowski space, 
 and using the relations Eqs.~(\ref{rel-3}), (\ref{rel-4}), and (\ref{rel-5}) of Appendix~\ref{set-relations}, it transforms into
\begin{equation}
\label{final-tadpole}
\Pi^{(t)}_T (p_0, {\bf p})= \frac{c_s^2p^2}{\rho_0} \int \frac{d^3q}{(2\pi)^3}\frac{d n_B}{d E_q}q^2
\left[\frac{1}{12}+\frac{\rho_0}{4c_s}\frac{\partial c_s}{\partial\rho_0}+\left(\frac{p_0}{c_sp}\right)^2\left(-\frac{3\rho_0}{4c_s}\frac{\partial c_s}{\partial\rho_0}+\frac{5\rho_0^2}{4c_s^2}\left(\frac{\partial c_s}{\partial\rho_0}\right)^2-\frac{\rho_0^2}{4c_s}\frac{\partial^2c_s}{\partial\rho_0^2}\right)\right]
\ .
\end{equation}

The lollypop diagram is computed in a similar way, and it gives
\begin{equation}
\Pi^{(l)}_T(p_0,p)=-\frac{\pi^2T^4p^2}{30c_s^3\rho_0}\frac{\rho_0}{c_s}\frac{\partial c_s}{\partial\rho_0}\left(1+\left(\frac{p_0}{c_sp}\right)^2\left(\frac{2\rho_0}{c_s}\frac{\partial c_s}{\partial\rho_0}-1\right)\right)
\end{equation}
where we have performed first the ${\bf q}$ integral in Eq.~(\ref{lolly}) and then the sum over Matsubara frequencies. Inverting the order of these two last operations gives an ill-defined result.
We have verified our computation by analyzing the diagram in configuration space, using a mixed representation of the phonon propagators, where no ambiguity is present.

Adding the different contributions obtained so far,
\begin{equation}
\Pi_T(p_0,{\bf p})  = \Pi^{(b)}_{1,T}(p_0,{\bf p})+	\Pi^{(b)}_{2,T}(p_0,{\bf p})+ \Pi^{(t)}_T (p_0,{\bf p}) + \Pi^{(l)}_{T}(p_0,{\bf p})\ ,
\end{equation}
and computing the  momentum integrals, one reaches the final result given in Eq.~(\ref{final-total}).

\begin{acknowledgments}
We thank M.~Mannarelli, D.T.~Son and J.~Soto for useful discussions.
This work has been supported by the Spanish grant
FPA2007-60275. M.A.E. has been also supported by MEC FPU (Spain).
\end{acknowledgments}


\begin{thebibliography}{99}
\bibitem{footnote}
In a recent publication \cite{Stachel} it has been
claimed that in Eq.~(\ref{NLO-Lagran}) there are structures proportional
to  $(D_t X)^2$ and  to $D_t X \nabla^2 \varphi$ which are absent.
However, using the equations of motion at LO, those pieces can be expressed in  terms
of those that are already written in Eq.~(\ref{NLO-Lagran}), so the Lagrangian proposed
by Son and Wingate is complete.
\end{thebibliography}
\end{document}